\documentclass[preprint,a4paper,3p,10pt,times,dvipdfm]{elsarticle}
\usepackage{graphics}
\usepackage{graphicx}
\usepackage{epsfig}
\usepackage{amssymb}
\usepackage{amsthm}
\usepackage{natbib}
\usepackage{latexsym}
\usepackage{mathrsfs}
\usepackage{amsmath}
\usepackage{amscd}
\usepackage{color}
\usepackage{verbatim}

\journal{Phys. Lett. A}

\newcommand{\fig}[1]{Fig.~\ref{#1}}
\newcommand{\eq}[1]{Eq.~(\ref{#1})}

\begin{document}

\begin{frontmatter}

\title{Footprints of sticky motion in the phase space of
  higher dimensional nonintegrable conservative systems}

\author[ufpr]{Cesar Manchein}
\ead{cmanchein@gmail.com}
\author[ufpr,mpipks]{Marcus W. Beims}
\ead{mbeims@fisica.ufpr.br}
\author[mpipks]{Jan M. Rost}

\address[ufpr]{Departamento de F\'\i sica, Universidade Federal do Paran\'a, 81531-980 
Curitiba, PR, Brazil}
\address[mpipks]{Max Planck Institute for the Physics of Complex
  Systems, N\"othnitzer Strasse 38, D-01187 Dresden, Germany}

\begin{abstract}
``Sticky'' motion in mixed phase space of high dimensional conservative 
systems is difficult to detect and to characterize. Its effect on 
quasi-regular motion is quantified here with four different measures, 
related to the distribution of the finite time Lyapunov exponents. We study 
systematically conservative maps from the uncoupled two-dimensional case 
up to coupled maps of dimension $20$. We find sticky motion in all
unstable directions above a threshold $K_{d}$ of the nonlinearity parameter 
$K$ for the high dimensional cases $d=10,20$. Moreover, as $K$ increases we 
can clearly identify the transition from  quasiregular to totally chaotic 
motion which occurs simultaneously in all unstable directions. Results show 
that all four statistical measures sensitively probe sticky motion in high 
dimensional systems.
\end{abstract}

\begin{keyword}
Finite Time Lyapunov Exponent Distribution\sep Stickiness \sep Chaos 
\sep High Dimensions
\end{keyword}
\end{frontmatter}

%

\section{Introduction}
\label{Introduction}
\begin{figure}[!ht]
\begin{minipage}[!t]{0.61\textwidth}
\centering
\includegraphics[width=1.0\textwidth]{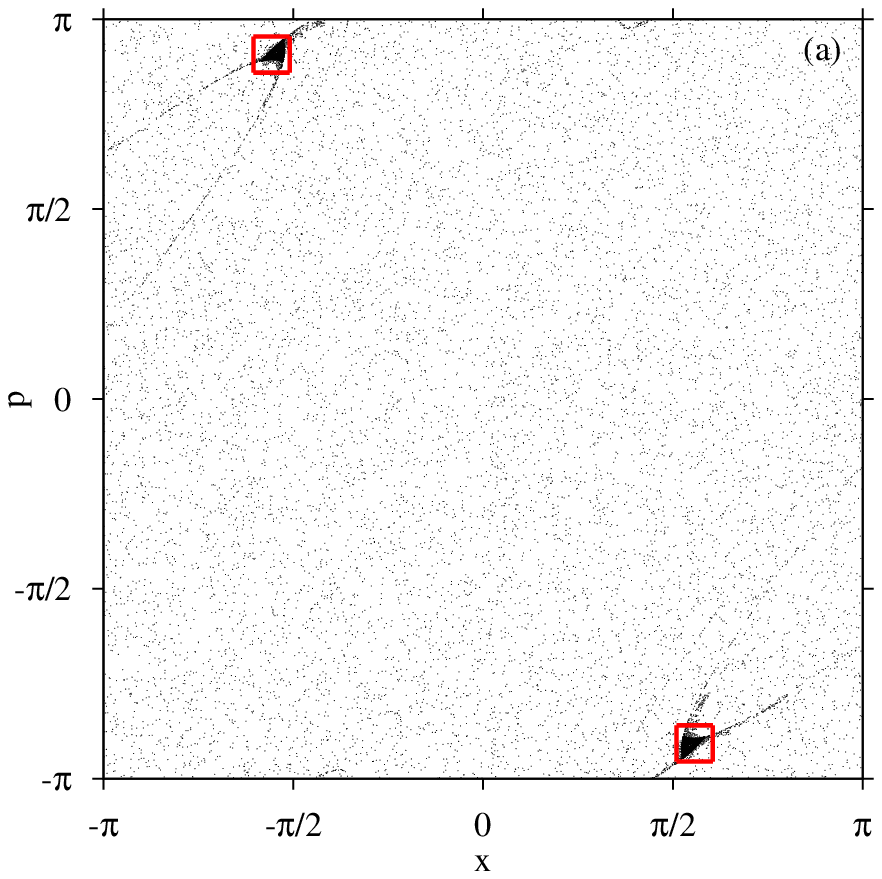}
\end{minipage}\hfill
\begin{minipage}[!t]{0.4\textwidth}
\includegraphics[width=0.8\textwidth]{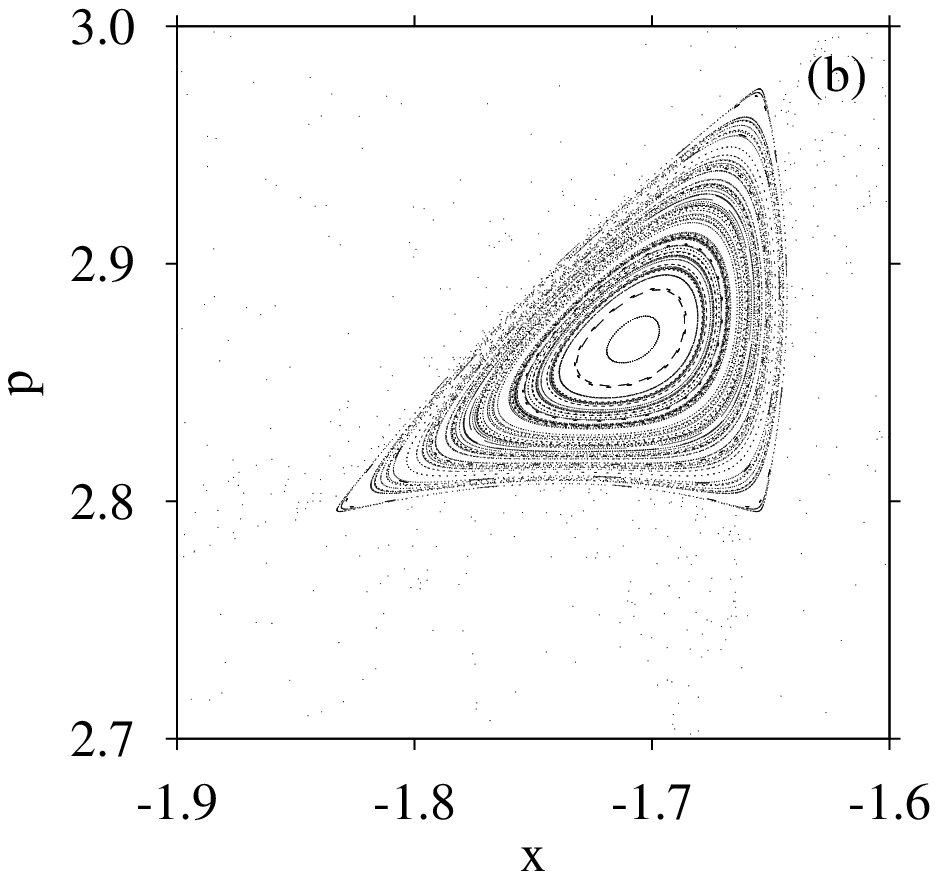}
\includegraphics[width=0.77\textwidth]{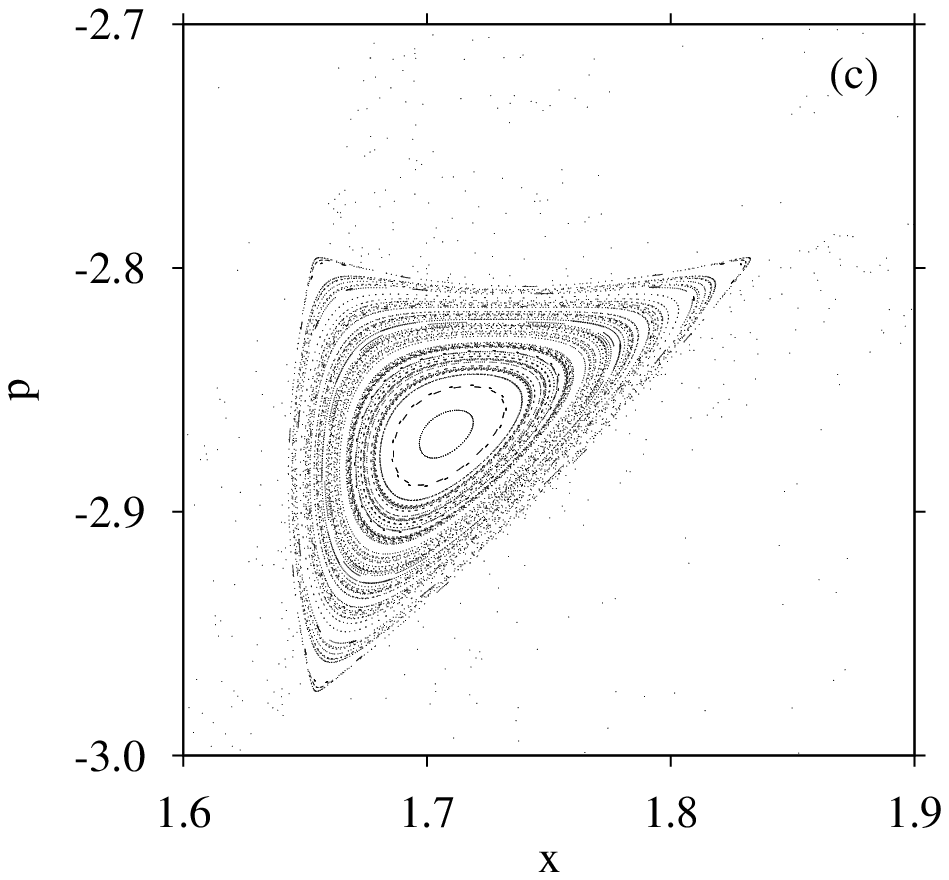}
\end{minipage}
\caption{Phase space for $K=5.79$. (b) and (c) are
         magnifications from the boxes shown in (a).}
\label{PSS1}
\end{figure}

The dynamics of conservative systems consists of different types of motion,
from regular over quasi-regular to chaotic, depending on the perturbation 
parameter.  Usually, in $d=2$ dimensional conservative systems
 regular islands (Kolmogorov-Arnold-Moser KAM tori) can break as the 
perturbation parameter increases, and chaotic trajectories may coexist with 
regular islands \cite{lichtenberg92}. This regime is called 
quasiregular. Chaotic trajectories may be trapped around stable islands 
creating  ``sticky'' motion (for a review see~\cite{zas-book}). In such 
cases the dynamics in phase space can be very structured and the motion is 
non-ergodic. Analytical results for physical quantities in the quasiregular 
regime are rare and it is very desirable to obtain tools which quantify 
degrees of ergodicity and hyperbolicity, as for example the angle between  
the stable and unstable manifolds obtained via the covariant Lyapunov 
vectors \cite{lapeyre02,politi07}. Even for apparently totally chaotic 
systems, tiny regular islands may appear which induce sticky motion as 
exemplified
in \fig{PSS1}a with the phase space of the $d=2$ dimensional kicked rotor
[Eqs.~(\ref{SM}] for the nonlinearity parameter $K=5.79$. The dynamics
looks totally chaotic -- yet, tiny regular islands appear near the
points $(x,p)\sim(4.6,2.9),(1.7,-2.9)$ (see boxes), if enough initial
conditions are sampled. Such islands become visible with sufficient
magnification (Figs.~\ref{PSS1}b-c). 

Hence, the question arises how these tiny regular islands
can be identified with  not more information than provided  by \fig{PSS1}a. 
In higher dimensions, the information equivalent to \fig{PSS1}a  must be 
cast in a different form since  Poincar\'e Surfaces of Section (PSS) cannot 
be plotted. In fact, for $d>2$ dimensional conservative systems, due to 
Arnold diffusion, the KAM surfaces do not divide the whole phase space into 
a set of closed volumes. Although the effect of
stickiness of $N$-dimensional invariant tori has been analyzed for the 
asymptotic decay of correlations and anomalous diffusion \cite{altmannEPL07},
to the best of our knowledge, the behavior of sticky trajectories in 
higher-dimensional systems has never been quantified and explored 
systematically. 

In the following we present extensive numerical simulations to detect and 
quantify  sticky motion in the phase space of quasiregular and chaotic 
conservative systems with up to 20 dimensions. To fulfill this 
task we analyze small changes in the finite time Lyapunov exponents (FTLE) 
distribution using four quantities: the variance $\sigma$, the skewness 
$\kappa_3$, the kurtosis $\kappa_4$ and  ${\cal P}_{\Lambda}$, the normalized 
number of occurencies  of the most probable finite time Lyapunov exponent. 
The first three quantities were analyzed in the two-dimensional standard map 
for the distribution of the finite time stability exponents to detect islands 
\cite{steven07} and to study the fluctuations of the mean FTLE for specific 
trajectories and the corresponding Kolmogorov like entropy
\cite{froeschle93}.

The FTLEs are the average of local divergence of trajectories and for ergodic 
systems they converge to the LE obtained in the limit of infinite times, which 
is independent of initial 
conditions~\cite{oseledec68}. In such cases the FTLE distribution has
Gaussian form (normal distribution) \cite{politi88,grebogi,falcioni}. However, 
in quasiregular systems the chaotic trajectories may visit the neighborhood 
of regular islands, being trapped there for a while, and the FTLEs do not 
converge quite well. Such small changes in the FTLEs distribution can be 
used for many purposes: to obtain information about algebraic or exponential 
stretching in phase space \cite{badii89,ott90}, to quantify ergodicity 
\cite{berry92,berry93}, to study stretching processes in chaotic flows 
\cite{wiggins93}, to detect the bimodal \cite{lopes05} or multimodal
 \cite{feudel07} behavior of the local FTLEs, to analyze the time correlations
\cite{kantz85,kantz87,manchein09} and Poincar\'e recurrence \cite{manchein09} 
in coupled maps, to detect the sticky motion  \cite{cesar1,hercules08} and 
the arise of regular islands \cite{cesar2} for interacting particles in 
one- and two-dimensional billiards  via ${\cal P}_{\Lambda}$ and, as we will
see here, 
to detect and quantify  sticky motion in higher-dimensional nonintegrable 
conservative systems. In addition, FTLEs have been used in dissipative systems
to distinguish between strange chaotic and nonchaotic attractors (see 
\cite{wang04,prasad01,feudel95,grebogi84} for some examples). In these
cases the trajectory switches between long times on the periodic 
attractor (negative FTLE) and smaller times on the chaotic saddle with 
positive FTLEs. This is quite the opposite of what is observed in 
conservative chaotic systems, where the trajectory { evolves for 
longer times in the chaotic region of the phase space} and smaller times near 
the regular island.

In contrast to the large literature related to the analysis of sticky 
motion via the distribution of the {\it local} FTLEs for some specific 
trajectories { and as a function of time}, in this work we 
analyze the FTLEs distribution  {\it after} $n=10^7$ iteration of the 
map and over $10^4$ initial conditions distributed uniformly in phase 
space. Our investigations show that even for such large 
times the FTLEs still depend on initial conditions when sticky motion is 
present. { Such times are large enough to give a good convergence 
of the FTLEs and to detect the sticky motion.
If the iteration times increase too much ($\gtrsim 10^{8}$), effects 
of sticky motion on the distribution start to dissapear since for 
infinite times the distribution approaches the $\delta-$function in 
chaotic systems. }

The paper is organized as follows. In Section \ref{model} we describe the 
properties of the variance  $\sigma$ and  the distribution of the most 
probable Lyapunov exponent ${\cal P}_{\Lambda}$ for a two-dimensional 
conservative system (the standard map) and compare them with higher 
cummulants  ($\kappa_{3}$ and $\kappa_4$). In Section \ref{coupled} we couple 
$N$ { conservative} maps and by  increasing  $N=2,3,5,10$ we analyze 
carefully the 
changes observed in the the quantities  ${\cal P}_{\Lambda}$, $\kappa_3$ and 
$\kappa_4$. Section \ref{conclusions} summarizes the main results of 
the paper.

\section{The standard map and the FTLE distributions}
\label{model}
The standard map is an ideal model for a conservative chaotic system
to study the behavior of FTLE distributions, specially if one is interested 
in an extension to higher dimensions. The original  
$d=2$ dimensional map $M$ was proposed by Chirikov \cite{chirikov79}
and is defined by 
\begin{eqnarray}
  M:\left\{
\begin{array}{ll}
  p_{n+1} = p_n + K\sin (x_n) & \qquad \mathrm{mod}\ 2\pi, \\
  x_{n+1} = x_n + p_{n+1}      & \qquad \mathrm{mod}\ 2\pi,
\end{array}
\right.
\label{SM}
\end{eqnarray}
where $p_n$ and $x_n$ are the dynamical variables of the system 
after the $n$th iteration and $K$ is the nonlinear parameter. This map 
locally approximates more general nonlinear mappings \cite{lichtenberg92}. 
The FTLE spectrum is determined { numerically} for a conservative 
system using { Bennetin's algorithm \cite{benettin80,wolf85}
which includes the Gram-Schmidt re-orthonormalization procedure.
The sum of all FTLEs is zero and they come in pairs $\pm \Lambda^{(i)}_{n}$ 
($i$ is for the $i$th unstable direction)}. Hence, it is sufficient to analyze 
the $\Lambda^{(i)}_{n}>0$. The FTLEs can be defined by 
\begin{equation}
  \Lambda_n^i({\bf x}) = \frac{1}{n} \ln \parallel 
  M({\bf x},n){\bf e}_i({\bf x}) \parallel,
\label{lambda}
\end{equation} 
with the discrete iterations $n$. $M({\bf x},n)=J({\bf x}_{n-1})
\cdots J({\bf x}_1)J({\bf x}_0)$,  and $J({\bf
  x}_n)~(n=0,\ldots,n-1)$ are the Jacobian matrices determined at
each iteration $n$. ${\bf e}_i({\bf x})$ represents the unit vector in the
$i$th unstable direction. These vectors are re-orthonormalized at each 
iteration and the FTLE spectrum is obtained using Eq.~(\ref{lambda}).
The usual LE is obtained from 
$\Lambda^i=\lim_{n\to\infty}  \Lambda_n^i({\bf x})$ which
is independent of the initial conditions. 
\begin{figure}[htb]
\centering
 \includegraphics*[width=8.2cm,angle=0]{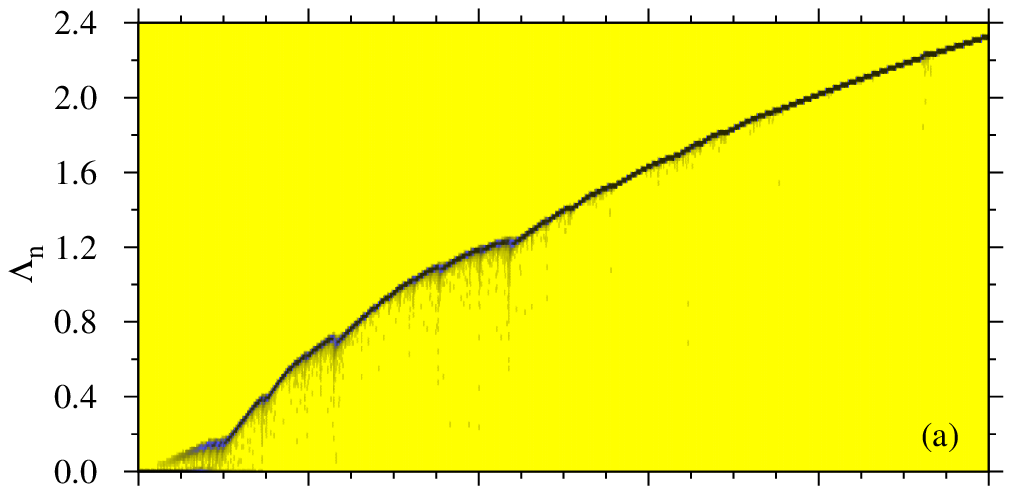}
 \includegraphics*[width=8.5cm,angle=0]{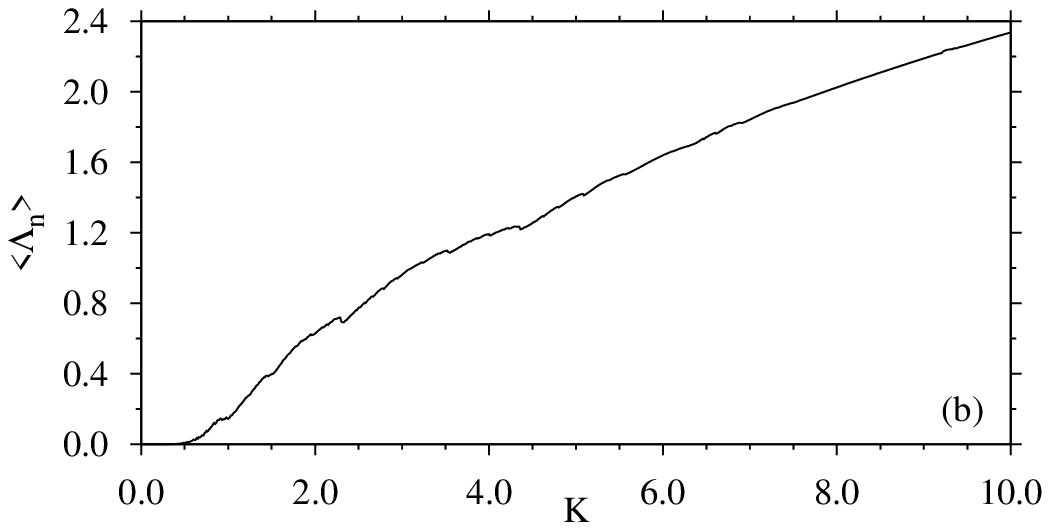}
\caption{(a) (Color online) Finite-time distribution of the positive Lyapunov 
 exponent $P(\Lambda_{n},K)$ calculated over $400$ trajectories up to 
  $n = 10^7$ iterations. With increasing $P(\Lambda_{n},K)$ the color changes 
 from light to dark (white over yellow and blue to black). (b) The corresponding
  mean value $\left<\Lambda_n\right>$.}
\label{MP}
\end{figure}

Figure \ref{MP}a shows the FTLEs distributions $P(\Lambda_{n},K)$ 
as a function of the nonlinearity parameter $K$ for $400$ 
initial conditions and  up to $n=10^7$ iterations. For each initial 
condition we 
reckon one FTLE $\Lambda_n$, which is the time average of all local FTLE 
along the trajectory. In Fig.~\ref{MP}a
the color changes from light to dark (white over yellow and blue to 
black) with increasing $P(\Lambda_{n},K)$.

The gray points below the main curve are related to FTLEs from chaotic 
trajectories which were trapped for a while close to regular islands. For 
these trapped chaotic trajectories (or ``sticky'' motion) the {\it local} FTLE 
decreases and consequently the final $\Lambda_n$ also decreases. In such cases 
the FTLE distribution is not Gaussian  anymore but contains interesting 
information about the dynamics. Figure \ref{MP}a also reveals that for some 
specific values of $K$ the amount of sticky trajectories changes, which will 
be quantified and characterized next. Note, that such
information is not provided by the mean FTLE shown in  Fig.~\ref{MP}b.
\subsection{Quantitative characterization of the distributions}
\label{quantity}
Clearly,  anomalies in the FTLE distribution reflect sticky motion - however, 
the entire distribution is not a convenient tool to identify and measure 
sticky motion.  Instead, we will analyze the following four measures:

\begin{description}
\item[a.)]  The number of occurencies of the most probable FTLEs 
$P(\Lambda_n^p,K)\equiv {\cal P}_{\Lambda}(K)$ which informs how 
many initial conditions (normalized) lead to the mode of the
distribution \cite{cesar1,hercules08,cesar2}.
${\cal P}_{\Lambda}(K)=1$ means that for all initial 
conditions the FTLEs are equal (within the precision $10^{-3}$). Therefore we 
expect ${\cal P}_{\Lambda}(K)=1$ for a totally chaotic (or totally regular) 
system and large times. 
In cases the sticky motion appears or if the normal distribution did not 
converge to the given precision, ${\cal P}_{\Lambda}(K)<1$. {
}

\item[b.)] The variance defined by 
\begin{equation}
\tilde\sigma={\left<(\Lambda_n-\left<\Lambda_n\right>)^2 \right>},
\end{equation}
which is the second cummulant of the FTLE distribution. It should increase 
for sticky motion. We will analyze the properties of the
relative variance $\sigma=\tilde\sigma/\left<\Lambda_n\right>^2$ which
is more appropriate for higher-dimensional systems. In such cases it is 
possible to detect small differences relative to each unstable direction.

\item[c.)] The skewness is defined by 

\begin{equation}
\kappa_3= \frac{\left<(\Lambda_n-\left<\Lambda_n\right>)^3\right>}
{\tilde\sigma^{3/2}}.
\end{equation}
It is the third cummulant of the FTLE distribution and detects the 
asymmetry of the distribution around its mean value. For $\kappa_3=0$ we 
have the regular distribution. Since sticky motion usually reduces 
the FTLE, the asymmetry of the distribution leads to $\kappa_3<0$.

\item[d.)] The kurtosis is defined by 
\begin{equation}
\kappa_4= \frac{\left<(\Lambda_n-\left<\Lambda_n\right>)^4\right>}
{\tilde\sigma^2}-3,
\end{equation}
and detects the shape of the distribution.
When $\kappa_4>0$ indicates that  the distribution is flatter than the
regular distribution ($\kappa_4=0$),
while $\kappa_4<0$ reveals a sharper distribution than the normal one.
\end{description}
These four quantities will be used to characterize the degree of sticky
motion in the transition from low- to higher-dimensional systems. Altogether 
they should give all relevant informations which can be extracted 
from the distributions.
\subsection{The two-dimensional case ($d=2$)}
\label{distrib1D}
To compare with the higher dimensional cases to be discussed later we summarize 
here $d=2$ results for the FTLEs distributions of the standard map in a
slightly different $K$ regime than considered by \cite{steven07}.
\begin{figure}[htb]
\centering
 \includegraphics*[width=9.0cm,angle=0]{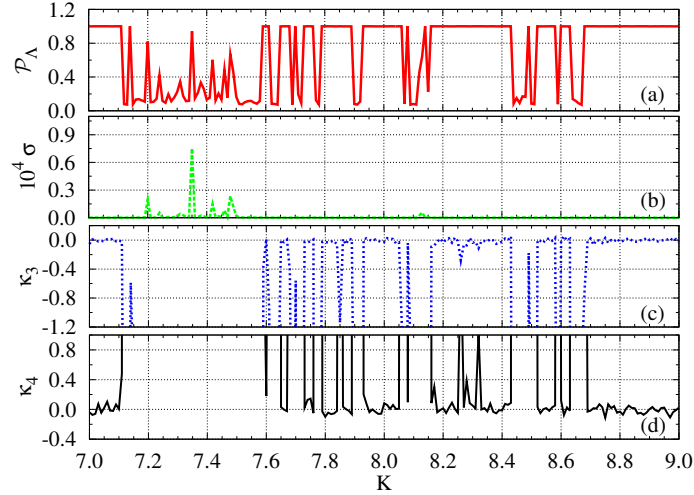}
\caption{(Color online) Deviations from the mean Lyapunov exponent in the 
interval $K=(7.0,9.0)$ 
for the standard map measured with (a) ${\cal P}_{\Lambda}(K)$, (b) $\sigma(K)$, (c) 
$\kappa_3(K)$ and (d) $\kappa_4(K)$.}
\label{N1}
\end{figure}
For all simulations we have used $201$ 
values of $K$ in the interval $K=(7.0,9.0)$ (stepsize $\Delta K = 0.01$).
As can be seen from  \fig{N1}a
${\cal P}_{\Lambda}$ has many minima in the interval $K=(7.0,9.0)$. For all 
minima of ${\cal P}_{\Lambda}$ a certain amount of sticky motion is expected. 
This is indeed true, as we have checked constructing 
the PSS (not shown) for some values of $K$ where minima in ${\cal P}_{\Lambda}$ 
appear and for all of them we found sticky motion. Comparing ${\cal P}_{\Lambda}$ 
with Fig.~\ref{N1}b we see that $\sigma$ has maxima only for values
of $K$ in the interval  $(7.15,7.55)$ and close to $8.12$, i.e., the variance 
does not detect all the sticky regions. However, skewness (\fig{N1}c) and kurtosis  
(\fig{N1}d) detect sticky motion as identified by ${\cal P}_{\Lambda}$  in  
(\fig{N1}a). Both parameters behave as expected. In
Fig.~\ref{N1}c we plotted the skewness $\kappa_3$ as a function of $K$. 
For almost all minima of ${\cal P}_{\Lambda}$ we observe that the skewness 
(\fig{N1}c) 
is negative. This implies that the sticky motion induces a left asymmetry in
the FTLEs distribution. In other words, the tail on the left side of the 
distribution is larger than the tail on the right side. For almost 
all minima of ${\cal P}_{\Lambda}$ the kurtosis $\kappa_4$ (\fig{N1}d) has a 
maximum. This is expected since a small ${\cal P}_{\Lambda}$ implies widely 
distributed $\Lambda_{n}$  and consequently a relatively flat distribution 
(large $\kappa_4$).

Obviously, in the two-dimensional case ${\cal P}_{\Lambda}$, and the cummulants 
$\kappa_3$ and $\kappa_4$ are sensitive and consistent detectors  of sticky 
phase space trajectories, with the latter two being more sensitive, while 
${\cal P}_{\Lambda}$ is less sensitive, but gives additional information (see, 
e.g., the interval $K=(7.2,7.5)$), which will become 
more obvious in the higher dimensional cases.

\section{Coupled {conservative} maps}
\label{coupled}
Coupled { conservative} maps are convenient systems to 
investigate  sticky motion in higher dimensional 
phase space systematically.
The equations for $N$ coupled { conservative} maps read
\begin{eqnarray}
  S:\left\{
\begin{array}{l}
   p_{n+1}^{(k)} = p_n^{(k)} + K\,\sin(x_n^{(k+1)}-x_n^{(k)}), \\
\\
  x_{n+1}^{(k)} = x_n^{(k)} + p_{n+1}^{(k)},
\end{array}
\right.
\label{CoupMaps}
\end{eqnarray}
where $k=1,\ldots,N$. We consider periodic boundary conditions
$p^{(N+1)}=p^{(1)},~x^{(N+1)}=x^{(1)}$ and unidirectional next 
nearest neighbor coupling. In \eq{CoupMaps} we  must use modulus $2\pi$. 
{ Translational invariance occurs in the $x$--coordinate and the 
Ly\-a\-punov spectrum always possesses two null exponents since two 
eigenvalues from the Jacobian of the coupled maps are equal to $1$. For 
$N=2$ our model can be written has a time-one map of a delta-function kicked 
Hamiltonian flow and the translational symmetry implies, via Noethers theorem, 
that the total momentum  $P_2= p_n^{(1)}+ p_n^{(2)}$ is a constant of motion.  
However, for $N>2$ our model cannot be written as a time-one map of a
Hamiltonian flow and the total momentum $P_N= \sum_k p_n^{(k)}$ is not a
constant of motion anymore (Noethers theorem is applied for Hamiltonian 
systems). In all numerical simulations we checked that the sum of all
LEs is close to  $10^{-14}$ and that their come in pairs. 

In addition, the condition $x_n^{(k+1)}-x_n^{(k)}=n\pi$, for $n=0,1,2\ldots$ 
is equivalent to the trivial uncoupled maps case $K=0$, where the linear 
momentum $p_n^{(k)}$ of each site is a constant of motion. Very close to this
condition we can use the approximation 
$\sin(x_n^{(k+1)}-x_n^{(k)})\approx x_n^{(k+1)}-x_n^{(k)}$, and
the total momentum becomes an approximated conserved quantity for any $N$.
This will be nicely observed in the simulations when stickiness occurs.}

\subsection{$N=2\, (d=4)$:  The separable case}
\begin{figure}[htb]
\begin{center}
 \includegraphics*[width=9.0cm,angle=0]{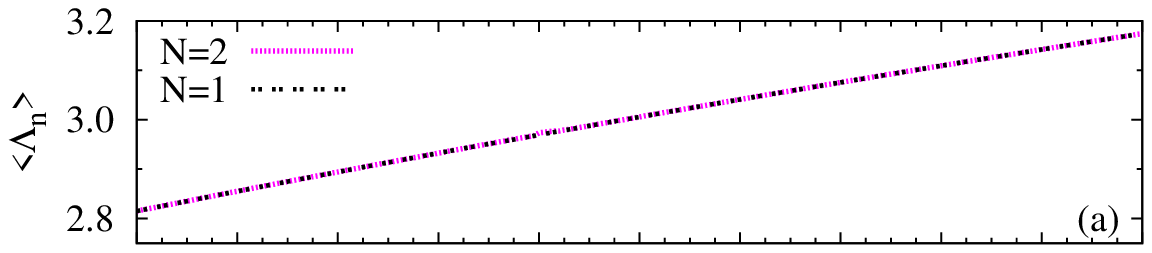}
 \includegraphics*[width=9.0cm,angle=0]{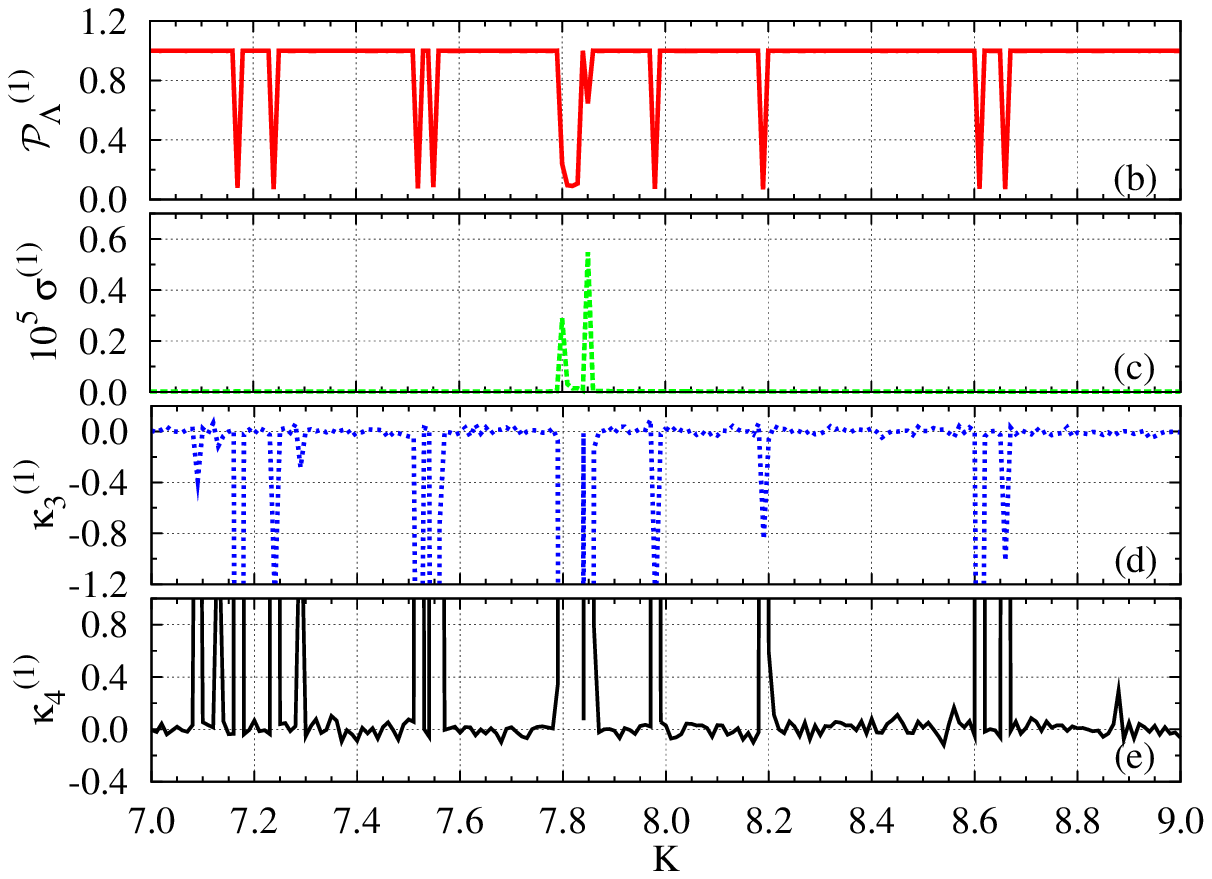}
\end{center}
\caption{(Color online) Comparison of (a) $\left<\Lambda_n\right>$ 
(the curve $N=1$ shows results for the standard map calculated 
at $2K=(14,18)$, see text) with (b) ${\cal P}_{\Lambda}$, (c)
$\sigma(K)$, (d) $\kappa_3(K)$ and (e) $\kappa_4(K)$ in the interval
$K=(7.0,9.0)$ and for $N=2$.} 
\label{N2}
\end{figure}

In this case the system \eq{CoupMaps} has  one positive, one negative and 
two zero  FTLEs. We show results related to the positive FTLE in \fig{N2}. 
Clearly, the mean value  $\left<\Lambda_n \right>$ increases with $K$ 
(\fig{N2}a). For $N=2$ the equations of $S$ (\ref{CoupMaps})
decouple in center of mass and relative coordinates. The map in the
relative coordinate is the standard map but with parameter
$2K$. Therefore, the curve in \fig{N2}a should match those obtained
for the standard map with parameter values  $2K=(14,18)$. To check
this we plotted $\left<\Lambda_n \right>$ in \fig{N2}a (dashed line)
for the standard map with values $2K=(14,18)$. Both curves are
identical. For all minima of ${\cal P}_{\Lambda}$ (\fig{N2}b) the skewness
$\kappa_3$ (\fig{N2}d) and the kurtosis  $\kappa_4$ (\fig{N2}e) have
also minima and maxima, respectively. At the corresponding values of
$K$  many trapped trajectories are expected. Since this case is the
standard map with larger nonlinearity parameter ($2K$), we do not
discuss further details.

\subsection{$N=3\, (d=6)$: The largest FTLE is most sensitive to 
sticky motion}
\begin{figure}[htb]
\centering
 \includegraphics*[width=9.0cm,angle=0]{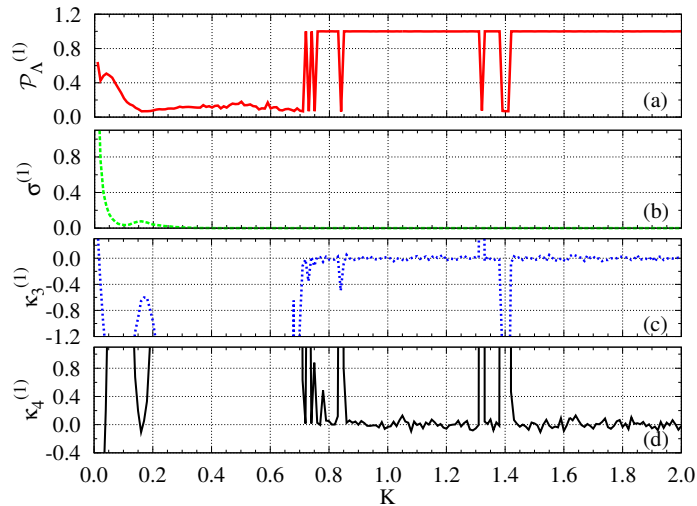}
\caption{(Color online) Comparison of (a) ${\cal P}_{\Lambda}^{(1)}(K)$ with (b) 
$\sigma^{(1)}(K)$,  (c) $\kappa_3^{(1)}(K)$ and (d) 
$\kappa_4^{(1)}(K)$ in the interval $K=(0.01,2.0)$ and $N=3$.}
\label{N3-1}
\end{figure}

\begin{figure}[htb]
\centering
 \includegraphics*[width=9.0cm,angle=0]{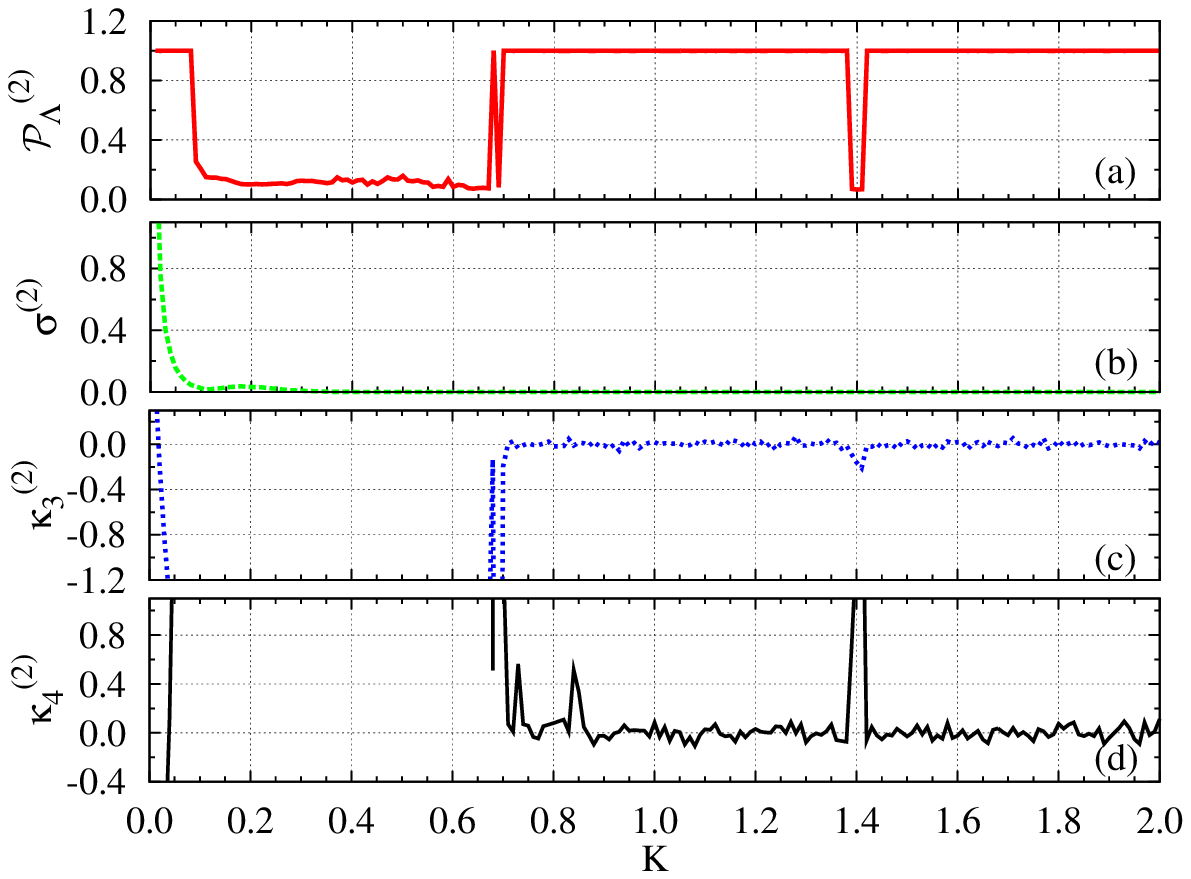}
\caption{(Color online) Comparison of (a) ${\cal P}_{\Lambda}^{(2)}(K)$ with (b) 
$\sigma^{(2)}(K)$, (c) $\kappa_3^{(2)}(K)$ and (d) 
$\kappa_4^{(2)}(K)$ in the interval $K=(0.01,2.0)$ and $N=3$.}
\label{N3-2}
\end{figure}

The case $N=3$ has six FTLEs, two positive, two negative and two zero. 
The analysis will be done for the distributions of the two positive 
FTLEs which are labeled such that  $\Lambda_n^{(1)}>\Lambda_n^{(2)}$. 
For $N=3$ the interval $(7.0,9.0)$  does not exhibit structures in the 
FTLE distributions pointing to sticky trajectories.
\begin{figure}[htb]
\centering
\includegraphics*[width=9.0cm,angle=0]{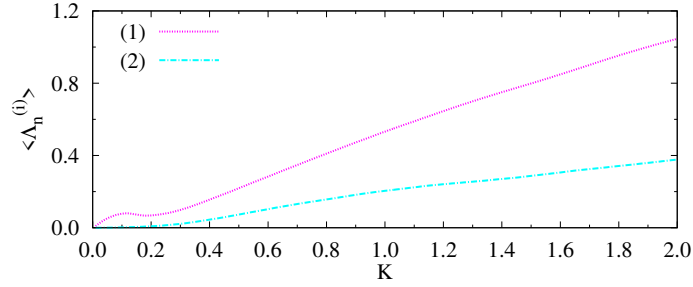}
\caption{(Color online) Mean values for the two positive FTLEs 
$\Lambda_n^{(1)}>\Lambda_n^{(2)}$ for $N=3$.}
\label{MN3}
\end{figure}
Hence, we analyze the interval $K=(0.01,2.0)$ which is 
close to the non-chaotic case $K=0$. Figure \ref{MN3} shows the mean values 
of the first and second FTLEs as a function of $K$. Both FTLEs increase with 
$K$, but $\Lambda_n^{(1)}$ increases faster and has a minimum close to 
$K\sim0.2$. Close to this value  $\Lambda_n^{(2)}$ starts to be finite. 
Figures \ref{N3-1} and \ref{N3-2} show the quantities 
(a) ${\cal P}_{\Lambda}^{(i)}(K)$, (b) $\sigma^{(i)}$, (d) $\kappa_3^{(i)}(K)$ and 
(d) $\kappa_4^{(i)}(K)$ as a function of $K$
for $\Lambda_n^{(i=1)}$ and  $\Lambda_n^{(i=2)}$, respectively.
By comparing these plots we see that both distributions detect the sticky
motion for $0.1\lesssim  K\lesssim0.7$. In this region the minimum  value of
$\kappa_3^{(i)}(K)$ is $\sim-50$ for $i=1$ and $-30$ for $i=2$.  The maximum 
value of $\kappa_4^{(i)}(K)$ is $4000$ for $i=1$ and $\sim2000$ for $i=2$.
For $K\lesssim0.1$ we observe in Fig.~\ref{N3-2} that 
${\cal P}_{\Lambda}^{(2)}(K)=1$, meaning that  all initial conditions lead to the
same  $\Lambda^{(2)}_n\sim 0.0$. However,  $\kappa_3^{(2)}(K)$ and 
$\kappa_4^{(2)}(K)$ show that the distribution around  
$\Lambda^{(2)}_n\sim 0.0$ is not Gaussian. By changing 
$K=0.01\rightarrow 0.1$ the distribution changes slightly from the right
asymmetric form [$\kappa_3^{(2)}(0.01)\sim 1$] to the left asymmetric form 
[$\kappa_3^{(2)}(0.1)\sim-3$]. The kurtosis is always flat 
($\kappa_4^{(2)}>-3$) but approaches a regular distribution
when $K\sim0.05$. Similar behavior is observed in Fig.~\ref{N3-1} for  
$\kappa_3^{(1)}(K)$ and $\kappa_4^{(1)}(K)$. The relative variance 
is shown to be relevant only at very low values of $K$ (for both FTLEs).

For $K\gtrsim0.7$ the distribution related to $\Lambda^{(1)}_n$ detects
the sticky motion at $K\sim 0.83, 1.40$ and  $K\sim 1.32$, while for
$\Lambda^{(2)}_n$ it detects $K\sim 1.40$ and $K\sim 0.83$ (but only 
$\kappa_4^{(2)}(K)$). This shows the intuitively
expected result that when the nonlinearity parameter increases the most 
unstable dimension (related to $\Lambda^{(1)}_n$) is more sensitive to the 
sticky motion. 

\subsection{$N=5\, (d=10)$: The smallest FTLE defines a critical 
nonlinearity $K_{d}$}
\begin{figure}[htb]
\centering
 \includegraphics*[width=8.7cm,angle=0]{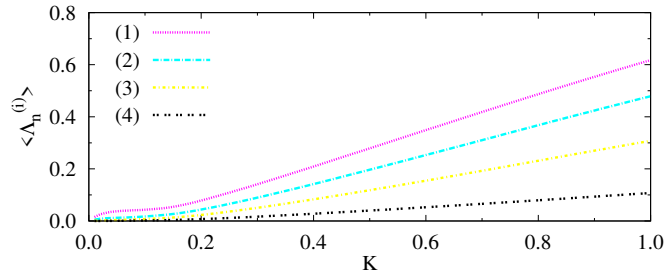}
\caption{(Color online) Mean values for the four positive FTLEs 
$\Lambda_n^{(1)}>\Lambda_n^{(2)}>\Lambda_n^{(3)}>\Lambda_n^{(4)}$ for $N=5$.}
\label{MN5}
\end{figure}
In Fig.~\ref{MN5} the mean values for the positive FTLEs 
$\Lambda_n^{(1)}>\Lambda_n^{(2)}>\Lambda_n^{(3)}>\Lambda_n^{(4)}$
are shown as a function of $K$. For smaller values of $K\lesssim0.15$,
$<\Lambda_n^{(3)}>$ and $<\Lambda_n^{(4)}>$ are almost zero, while 
$<\Lambda_n^{(1)}>$ increases initially very fast for $K=0.01\rightarrow0.05$ 
and then slowly for  $K=0.05\rightarrow0.15$. For $K$ above 
$0.2$ all mean FTLEs increase linearly with  different slopes size-wise 
ordered as the FTLEs themselves.
\begin{figure}[htb]
\centering
\includegraphics*[width=9cm,angle=0]{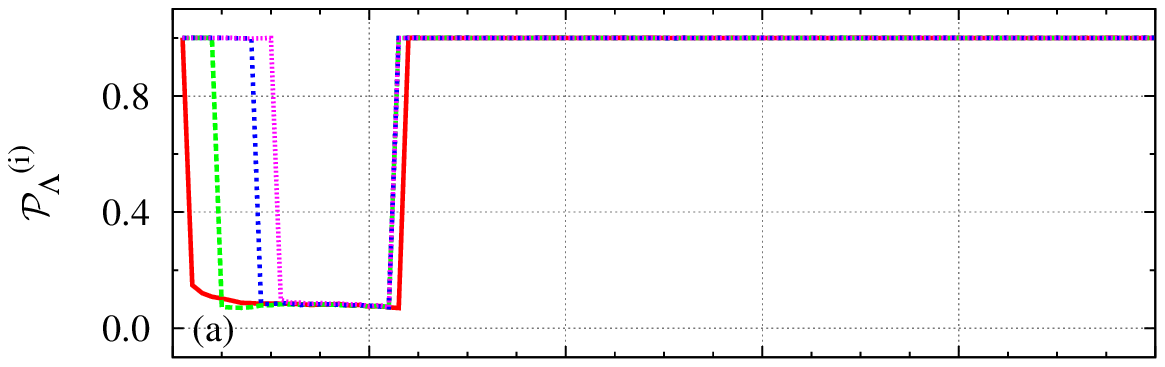}
\includegraphics*[width=9cm,angle=0]{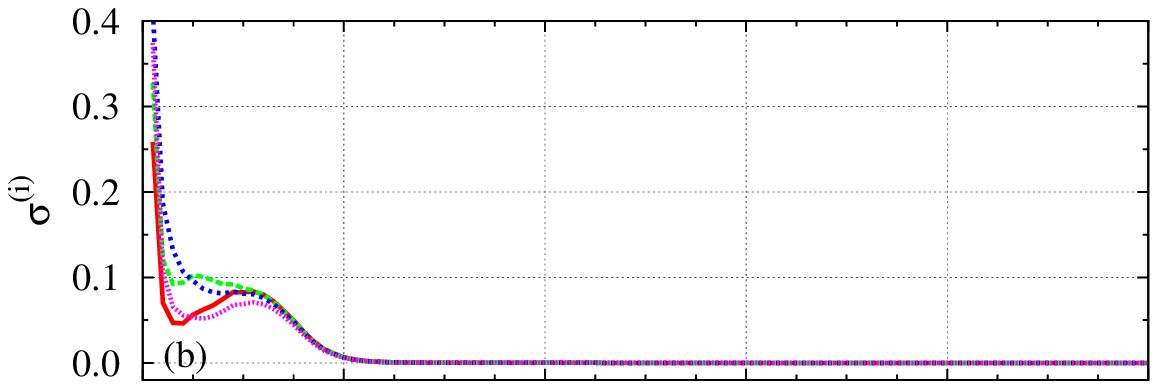}
\includegraphics*[width=9cm,angle=0]{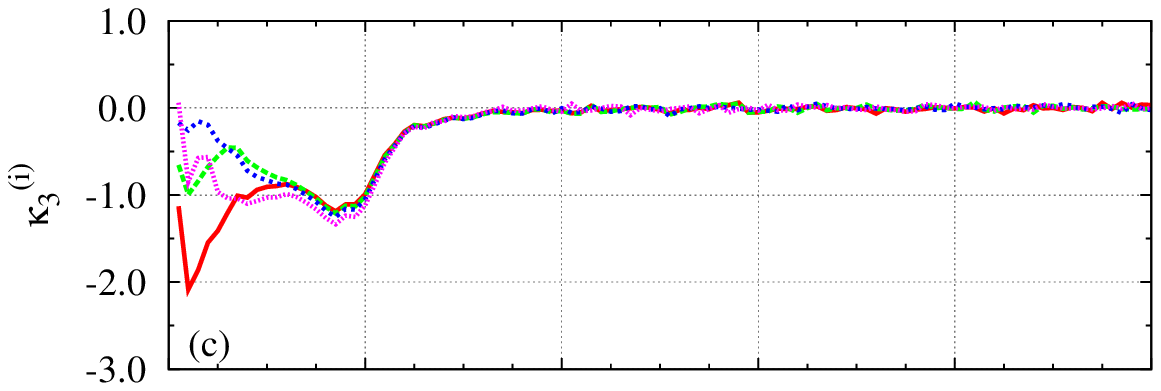}
\includegraphics*[width=9cm,angle=0]{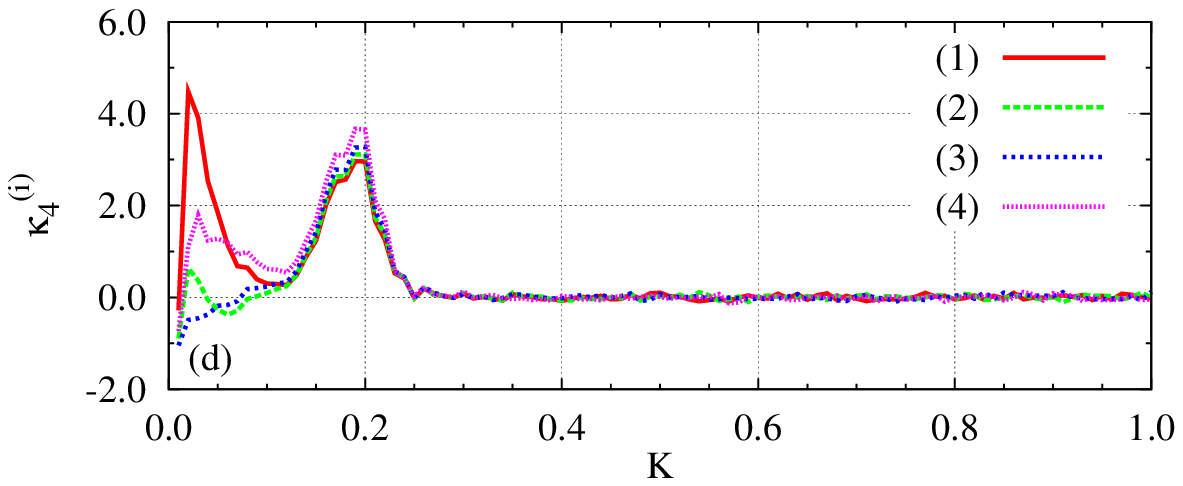}
\caption{(Color online) (a) ${\cal P}_{\Lambda}^{(i)}(K)$, (b)  
$\sigma^{(i)}(K)$, (c) $\kappa_3^{(i)}(K)$ and (d) $\kappa_4^{(i)}(K)$ in 
the interval $K=(0.01,1.0)$, $N=5$ and for all $i=1,2,3,4$.}
\label{N5}
\end{figure}

\begin{figure}[htb]
\centering
 \includegraphics*[width=9.0cm,angle=0]{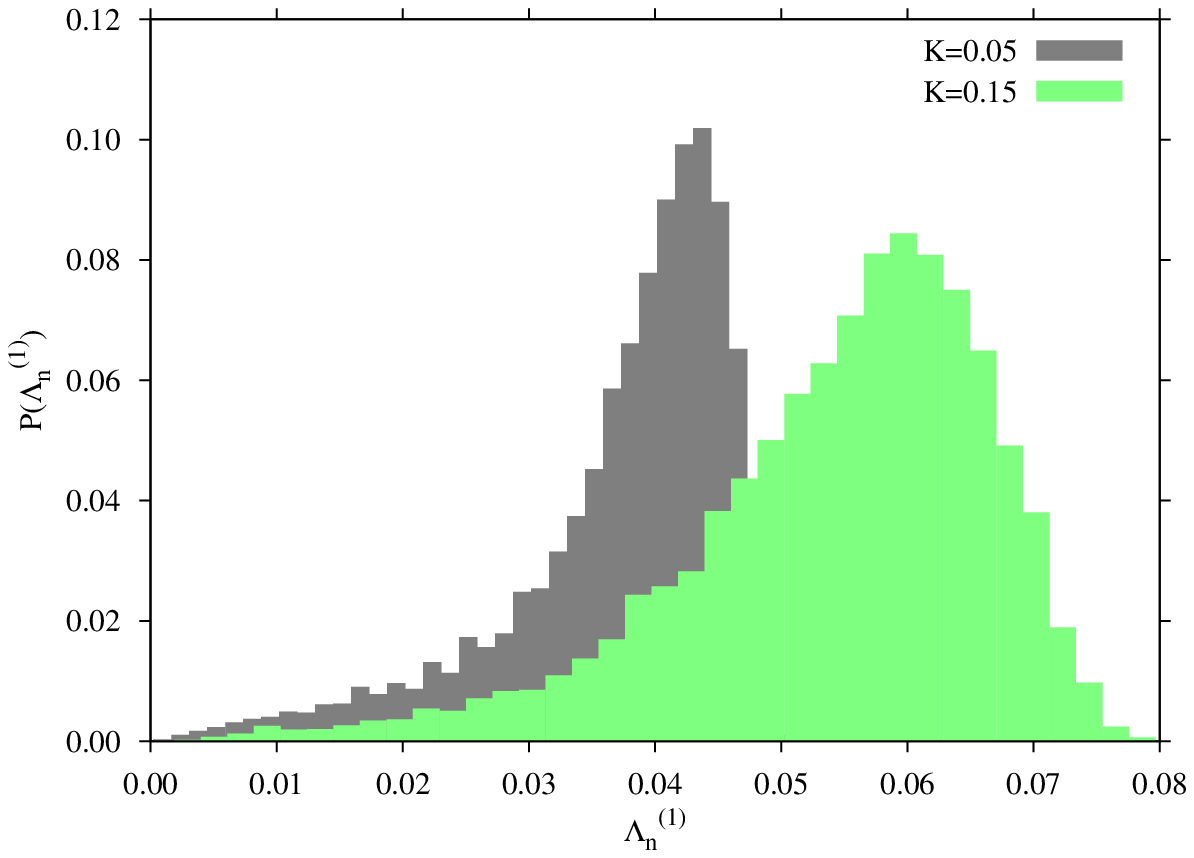}
\caption{(Color online) Distributions of the FTLEs for 
$K=0.05$ and $K=0.15$.}
\label{N5d}
\end{figure}

Figure \ref{N5} shows that for $K\gtrsim0.3$ 
the distributions from all FTLEs approach the normal (Gaussian)
form expected for a totally chaotic regime in contrast to the cases
$N=1,2,3$ discussed before, where  sticky motion also occurs for higher values 
of $K$. Therefore, there is a clear transition from the quasiregular to the 
totally chaotic dynamics at $K_c\sim0.3$ {\it simultaneously} in all unstable 
directions. On the other hand, all quantities detect the sticky motion for 
$K\lesssim0.25$. Interestingly, above a critical value $K_d\sim0.13$, the four 
indicators of sticky motion have a
similar qualitative behavior for all  FTLEs, i.~e., independently of the 
unstable directions. For example, the distributions for {\it all} FTLEs 
have a small left asymmetry ($\kappa_3\sim-0.75$) with no flatness 
($\kappa_4\sim0$) at $K=0.13$, and a left asymmetry ($\kappa_3\sim-1.50$) 
with flatness ($\kappa_4\sim3.0$) at $K=0.2$. 
{ Two exemplary 
distributions are shown in Fig.~\ref{N5d} for $K=0.05$ and $K=0.15$. Left 
asymmetry and flatness are evident.}
The critical value 
$K_d\sim0.13$ is  the point where ${\cal P}_{\Lambda}^{(4)}(K)$ detects 
the sticky motion from the  distribution of the last (the {\it smallest}) 
FTLE [see Fig.~\ref{N5}a]. The value  of $K_d$ can also be determined 
approximately from the onset of linear $K$-dependence 
of the mean FTLEs in \fig{MN5}.

\begin{figure}[htb]
\centering
\includegraphics*[width=10cm,angle=0]{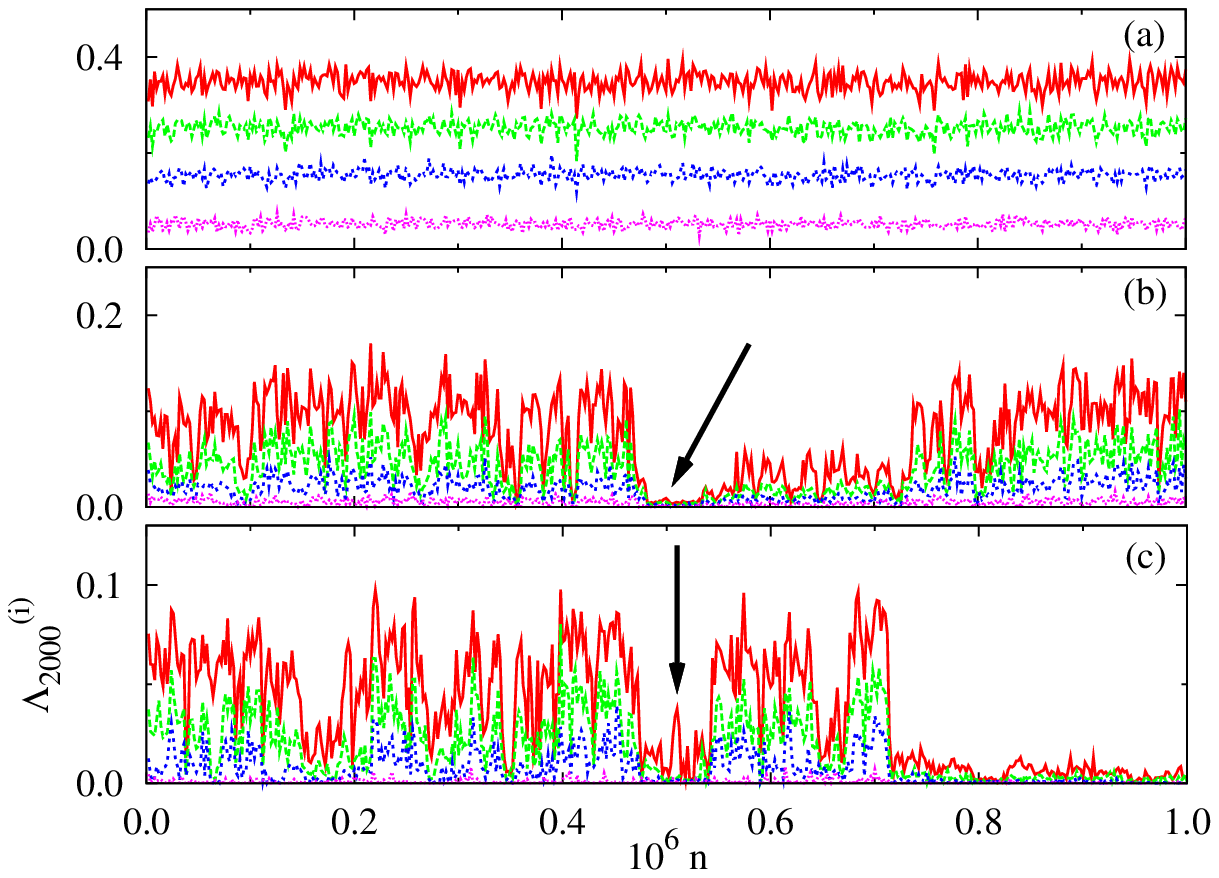}
\caption{(Color online) Time evolution of $\Lambda_n^{(i)}$ ($i=1,2,3,4$) 
for one exemplary trajectory with $N=5$ and (a)  $K=0.6>K_c$, (b) 
$K_d<K=0.15<K_c$, (c) $K=0.05<K_d$.}
\label{traj5}
\end{figure}

\begin{figure}[htb]
\centering
\includegraphics*[width=10cm,angle=0]{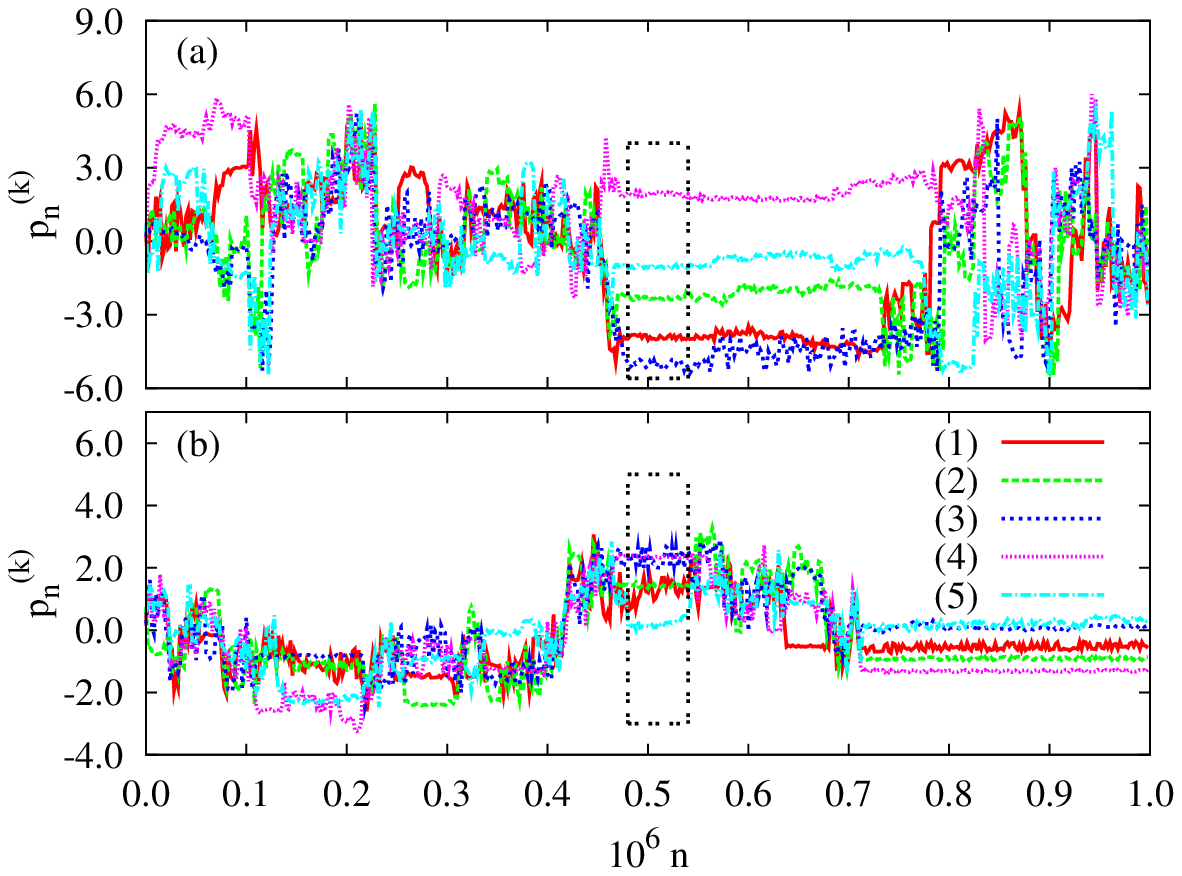}
\caption{(Color online) Time evolution of $p_n^{(k)}$ ($k=1,2,3,4,5$) 
for the trajectory from fig \ref{traj5}  (a) $K_d<K=0.15<K_c$, (b) 
$K=0.05<K_d$. Black boxes show the sticky region.}
\label{p5}
\end{figure}

These results show that the sticky motion induces 
differences between the distributions of  $\Lambda_n^{(i)}$ only for 
$K<K_d$. Since sticky motion is a consequence of regular structures in phase 
space, this suggests that there could exist independent invariant 
regions in phase-space. In the interval $K_d<K<K_c$ on the other hand, the 
sticky motion affects all unstable dimensions by the same amount leading to 
similar distributions for different FTLEs. Finally, in the totally chaotic 
regime $K>K_c$ 
all unstable directions are affected proportionally to each other. To explain 
this better we show in Fig.~\ref{traj5} the time evolution of 
$\Lambda_n^{(i)}$ ($i=1,2,3,4$) for one exemplary trajectory in 
the three distinct regions of $K$. Figure \ref{traj5}(a) shows the 
behavior of the FTLEs for $K=0.6>K_c$ and no sticky motion is 
observed since no FTLE goes to zero for some shorter time 
intervals. For the intermediate case $K_d<K=0.15<K_c$ we see in 
Fig.~\ref{traj5}(b) that for times around $0.5\times10^6$ (see arrow) 
{\it all} four FTLEs are zero, so the sticky motion occurs in all 
unstable directions and a common behavior is observed.
Physically what happens is that the linear momentum of {\it each} 
site becomes approximately constant during sticky motion. This is clearly 
shown in Fig.\ref{p5}(a), where $p_n^{(k)}\ (k=1,2,3,4,5)$ is plotted as
a function of $n$. At the sticky time $n\sim 0.5\times 10^6$ (see black box), 
$p_n^{(k)}$ is constant for {\it all} sites.
This agrees with the conjecture made by Froeschl\'e \cite{froeschle72} and
observed later in coupled maps by \cite{kantz87}, i.~e., that a conservative
system with $N\ge3$ has $N$ or one (the energy) constants of motion. In other 
words, following this conjecture when the sticky motion occurs all FTLEs 
should temporarily be very close to zero, as already observed. However, this 
changes 
when the region $K<K_d$ is analyzed, as can be observed in Fig.~\ref{traj5}(c) 
for times around $0.5\times10^6$ and $0.85\times10^6$ (see arrows). Only 
$\Lambda_n^{(1)}$ is larger and the remaining three FTLEs are very close to 
zero. In this case,  at the sticky time $n\sim 0.5\times 10^6$ (see black
box), the momenta $p_n^{(1)},p_n^{(3)}$ and $p_n^{(5)}$ 
are not conserved anymore. See Fig.~\ref{p5}(b).

\subsection{$N=10 \, (d=20)$}
The case $N=10$ has nine positive FTLEs. In this section we analyze again 
the interval $K=(0.01,1.0)$ but with $10^3$ initial conditions only.
\begin{figure}[htb]
\centering
 \includegraphics*[width=9cm,angle=0]{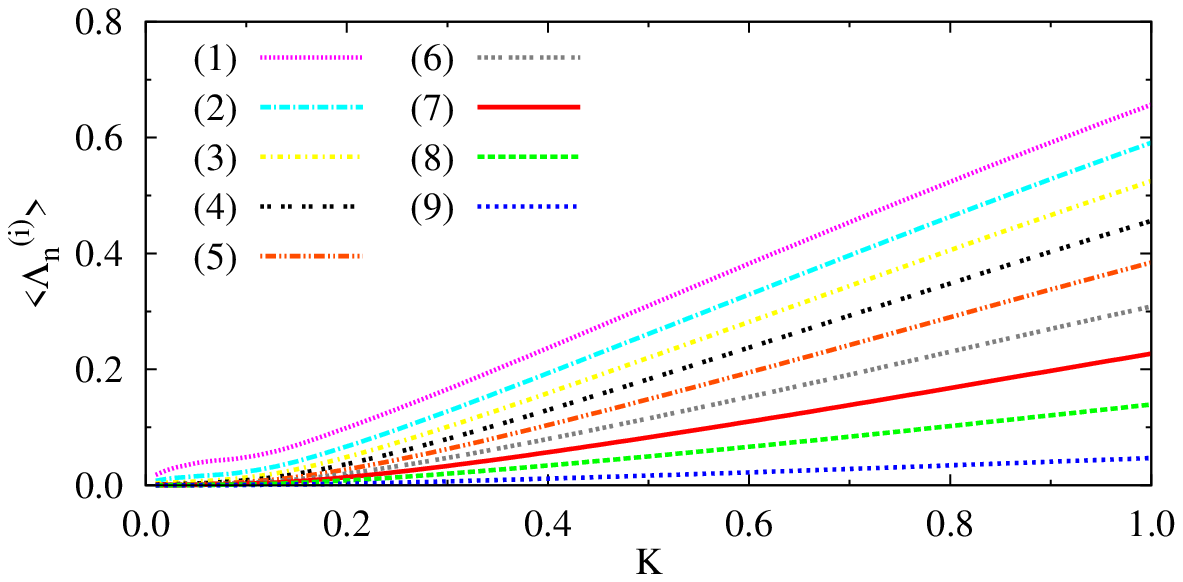}
\caption{(Color online) Mean values for the nine positive FTLEs 
$\Lambda_n^{(1)}>\Lambda_n^{(2)}>\ldots>\Lambda_n^{(9)}$, for $N=10$.}
\label{MN10}
\end{figure}
In Fig.~\ref{MN10} the mean values for the FTLEs are plotted as a function of 
$K$. For $K\gtrsim0.2$ all mean FTLEs increase linearly and the largest ones 
increase faster. For $K\lesssim0.2$ the $K$ dependence is different for 
different FTLEs. For $K\lesssim0.1$ almost all (but two $i=1,2$) are very
close to zero.
\begin{figure}[htb]
\centering
\includegraphics*[width=9cm,angle=0]{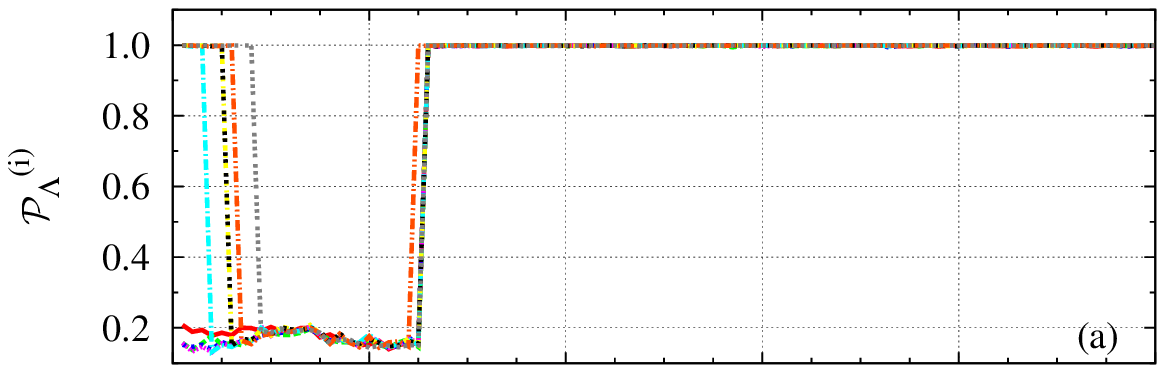}
\includegraphics*[width=9cm,angle=0]{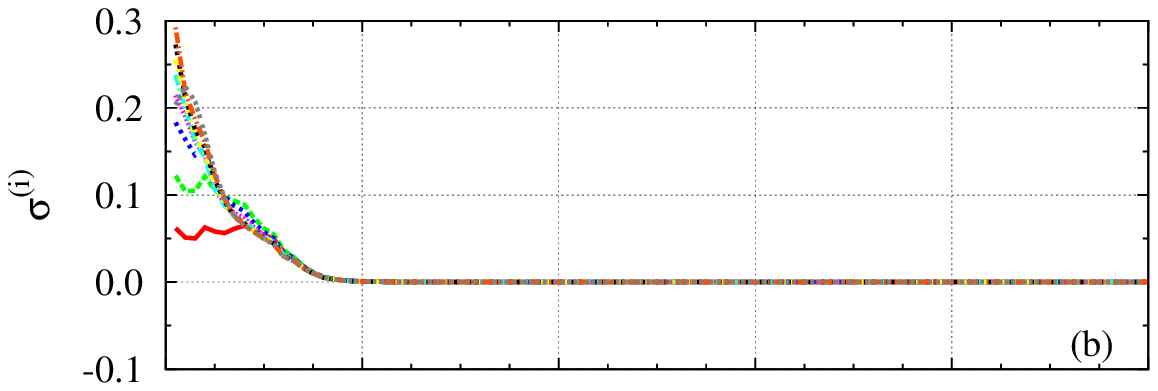}
\includegraphics*[width=9cm,angle=0]{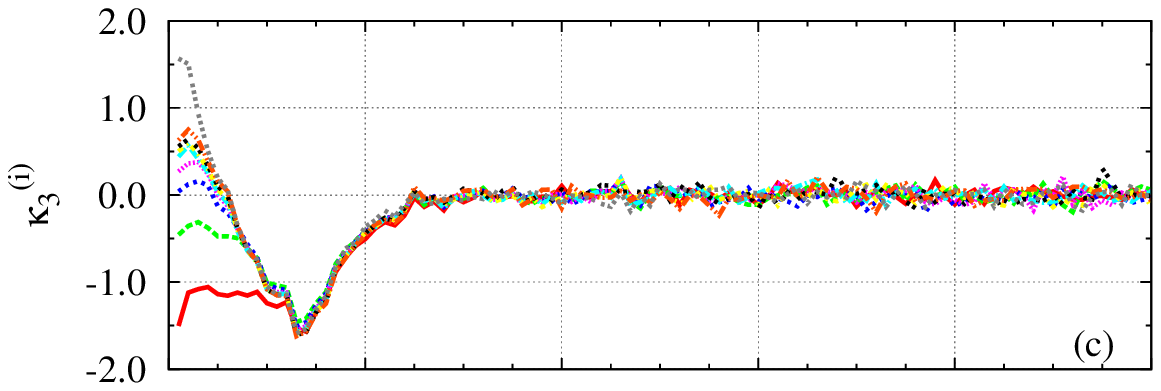}
\includegraphics*[width=9cm,angle=0]{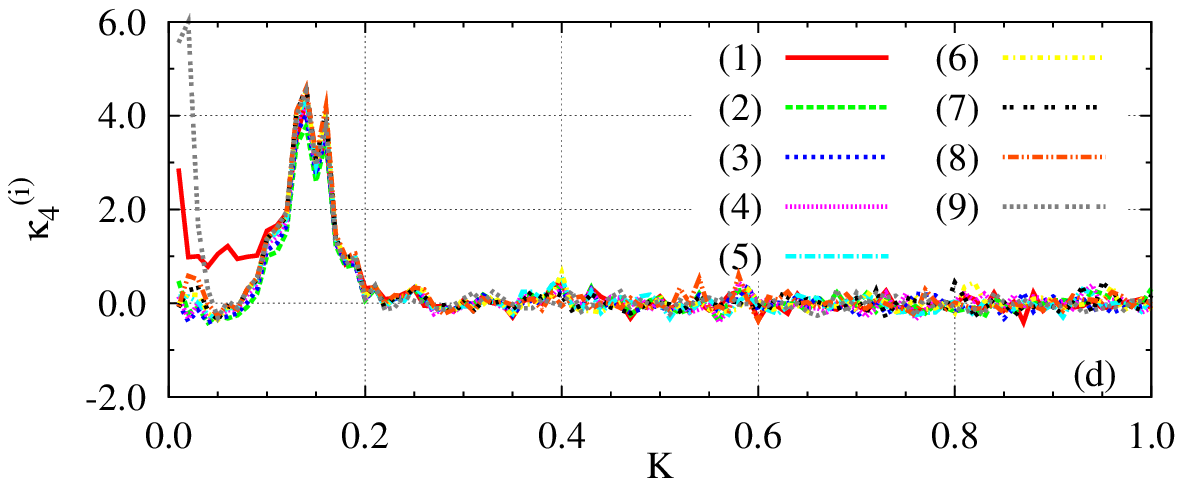}
\caption{(Color online) Quantities (a) ${\cal P}_{\Lambda}^{(i)}(K)$, (b) 
$\sigma^{(i)}(K)$, (c) $\kappa_3^{(i)}(K)$ and (d) $\kappa_4^{(i)}(K)$ 
as a function of $K=(0.01,1.0)$, for $i=1\rightarrow9$ and $N=10$.}
\label{N10}
\end{figure}
Figure \ref{N10} compares (a) ${\cal P}_{\Lambda}^{(i)}(K)$, (b) $\sigma^{(i)}(K)$, 
(c) $\kappa_3^{(i)}(K)$ and (d) $\kappa_4^{(i)}(K)$ in the interval 
$K=(0.01,1.0)$ for all positive FTLEs ($i=1\rightarrow9$).  For $K>K_c\sim0.28$
all quantities indicate that the distributions from all FTLEs approach the 
normal form expected for a totally chaotic regime. Therefore, again we have a 
clear transition from quasi-regular to totally chaotic dynamics at $K_c$ 
which occurs {\it simultaneously} in all unstable directions. For  
$K<K_c$ all quantities detect the sticky motion. Also here we
observe that for $K$ above the critical value $K_d\sim0.09$  the four 
quantities behave similarly, independent of the FTLEs.
The distributions for all FTLEs have a small left asymmetry 
($\kappa_3\sim-0.75$) with no flatness ($\kappa_4\sim0$) at $K=0.13$, 
and a left asymmetry ($\kappa_3\sim-1.50$) with flatness 
($\kappa_4\sim3.0$) at $K=0.2$. The critical value $K_d\sim0.09$
is again exactly the point where ${\cal P}_{\Lambda}^{(9)}(K)$ detects the sticky 
motion from the  distribution of the last (the {\it smallest}) FTLE 
[see Fig.~\ref{N5}(a)]. It can be observed that the distribution of
the largest FTLE is the last (when $K$ increases from $0.01$) to ``join''
the common behavior for $K_d\gtrsim0.09$ [please see 
Fig.~\ref{N10}(b)-(d)].
The value $K_d$ can also be determined approximately from the region 
in Fig.~\ref{MN10} where the $K$ dependence of the mean FTLEs is not linear.

\begin{figure}[htb] 
\centering
\includegraphics*[width=10cm,angle=0]{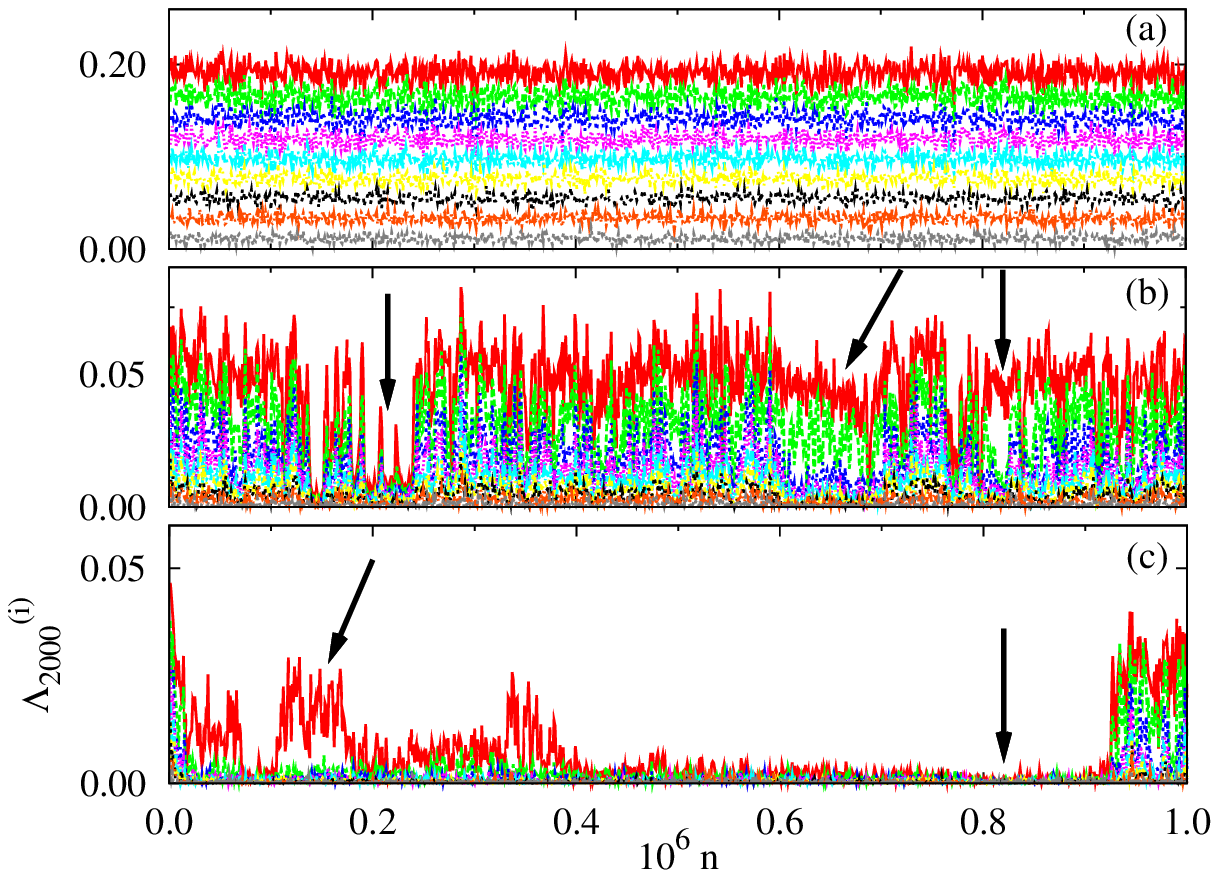}
\caption{(Color online) Time evolution of $\Lambda_n^{(i)}$ 
($i=1\rightarrow 9$) for one exemplary trajectory with $N=10$ and (a)  
$K=0.6>K_c$, (b) $K_d<K=0.15<K_c$, (c) $K=0.05<K_d$.}
\label{traj10}
\end{figure}

Figure \ref{traj10} shows the time evolution of  $\Lambda_n^{(i)}$ 
($i=1\rightarrow 9$) for one exemplary trajectory in the three distinct 
regions of $K$. In  \fig{traj10}(a) with $K=0.6>K_c$ no sticky motion 
is observed since no FTLE goes to zero for some shorter time 
intervals. For the intermediate case $K_d<K=0.15<K_c$ we see in 
Fig.~\ref{traj10}(b) the two different scenarios as conjectured. 
For times around $0.22\times10^6$ (see arrow) {\it all} nine FTLEs are 
close to zero indicating that the sticky motion occurs 
in all unstable directions.  For times around $0.8\times10^6$ (see arrow)  
on the other hand, eight FTLEs go to zero but the largest one remains 
positive.
This behavior is observed even more clearly in the region $K<K_d$, 
see Fig.~\ref{traj10}(c). For times in the interval 
$0.12\times10^6$ to $0.18\times10^6$ only  $\Lambda_n^{(1)}$ is large and 
the remaining eight FTLEs are very close to zero.

\section{Conclusions}
\label{conclusions}
It is well known that the dynamics of conservative systems consists of 
qualitatively different types of motion, from regular over quasi-regular to 
chaotic, depending on the perturbation parameter. It is difficult to identify
these types of motion in high-dimensional systems, especially in the
quasi-regular regime where the sticky motion can suppress  Arnold diffusion 
and distinct stochastic regions with different properties may appear 
\cite{vulpiani84,politi87}. Therefore, it is very desirable to develop tools 
which quantify the degree of sticky motion, hyperbolicity and 
ergodicity. Following ideas from \cite{steven07} the 
purpose of the present work is to quantify the degree 
of sticky motion in the phase space of higher-dimensional systems. To 
fulfill this task we have systematically analyzed small changes in the FTLE 
distributions using the skewness $\kappa_3$, which detects the asymmetry of 
the distribution, the kurtosis $\kappa_4$, which measures the flatness of 
the distribution and the normalized number of occurencies of the most 
probable finite time Lyapunov exponent ${\cal P}_{\Lambda}$, which indicates to 
which extent all trajectories started are under the influence of the same 
FTLE.

Using coupled { conservative} maps with nonlinearity parameter $K$ 
we show that the quantities  $\kappa_3(K),\kappa_4(K)$ and 
${\cal P}_{\Lambda}(K)$ are very sensitive to detect the sticky motion and 
capable to 
characterize the dynamics when the dimension $d$ of the system increases 
($d=2,4,6,10,20$).  { Si\-mu\-la\-tions were realized up to times 
$10^7$, which are large enough  to detect the sticky motion but not too large 
to destroy the sticky effect on the distribution, which for infinite times 
should approach the $\delta-$function. This should be valid for not too small 
values of the nonlinearity $K$. For very small values of $K$ and higher 
dimensions, the large amount of invariant structures make trajectories stay 
almost all times at the sticky motion. In such cases it is hard to say what 
times should be used in general.
} Besides detecting accurately all cases 
of sticky motion, we found the following additional interesting results: 
i) ($d=6$) the largest FTLE is most sensitive to 
the sticky motion; ii) ($d=10,20$) the smallest FTLE defines the critical 
nonlinearity parameter $K_d$, below which independent invariant regions in 
phase-space are expected and $\Lambda_n^{(i)}(K)$ behaves nonlinearly. For 
$K>K_c$ a clear transition from the quasiregular to the totally chaotic 
dynamics occurs 
{\it simultaneously} in all unstable directions. In the interval $K_d<K<K_c$
all quantities detect the presence of  sticky motion and  distributions for 
{\it different} FTLEs behave {\it similarly}, meaning that the sticky motion 
affects all unstable directions proportionally by the same amount and a kind 
of common behavior of the distributions is observed. 
Physically what happens is that the linear momentum of {\it all} 
sites is constant at the sticky motion.

Close-to-zero FTLEs play also an important role in other higher-dimensional 
system, for example in connection 
to hydrodynamic modes \cite{radons08,romero08} (among others) 
and the analysis of hyperbolicity \cite{steven07,viana08}. Results of the 
present work  contribute to categorize  the rich and 
complicated quasiregular dynamics of higher-dimensional conservative systems
and also to understand how nonlinearity is distributed along different 
unstable directions \cite{kantz85,janePD09}. 

\section{Acknowledgments}
The authors thank FINEP (under project CTINFRA-1) for financial
support. C.M.~also thanks CNPq for financial support and Prof. Giulio Casati 
for the hospitality by the Center for Nonlinear and Complex Systems at the 
Universit\`a dell'Insubria where part of this work was done.


\end{document}